\documentclass[%
 reprint,
superscriptaddress,
 amsmath,amssymb,
 aps,
floatfix,
]{revtex4-1}
\usepackage{graphicx}
\graphicspath{{images/}}
\usepackage{dcolumn}
\usepackage{bm}
\usepackage{hyperref}
\usepackage{xcolor}
\usepackage{xr}

\makeatletter
\newcommand*{\addFileDependency}[1]{
  \typeout{(#1)}
  \@addtofilelist{#1}
  \IfFileExists{#1}{}{\typeout{No file #1.}}
}
\makeatother

\begin{document}

\preprint{APS/123-QED}

\title{Novel Copper Fluoride Analogs of Cuprates}

\author{Nikita~Rybin}
\affiliation{Skolkovo Institute of Science and Technology, 30 Bolshoy Boulevard, bld.~1, Moscow 121205, Russia}
\author{Dmitry~Y.~Novoselov}
\affiliation{Skolkovo Institute of Science and Technology, 30 Bolshoy Boulevard, bld.~1, Moscow 121205, Russia}
\affiliation{M.~N.~Mikheev Institute of Metal Physics of Ural Branch of Russian Academy of Sciences, 18 S. Kovalevskaya St., Yekaterinburg 620137, Russia}
\affiliation{Department of Theoretical Physics and Applied Mathematics, Ural Federal University, 19 Mira St., Yekaterinburg 620002, Russia}
\author{Dmitry~M.~Korotin}
\affiliation{Skolkovo Institute of Science and Technology, 30 Bolshoy Boulevard, bld.~1, Moscow 121205, Russia}
\affiliation{M.~N.~Mikheev Institute of Metal Physics of Ural Branch of Russian Academy of Sciences, 18 S. Kovalevskaya St., Yekaterinburg 620137, Russia}
\author{Vladimir~I.~Anisimov}
\affiliation{Skolkovo Institute of Science and Technology, 30 Bolshoy Boulevard, bld.~1, Moscow 121205, Russia}
\affiliation{M.~N.~Mikheev Institute of Metal Physics of Ural Branch of Russian Academy of Sciences, 18 S. Kovalevskaya St., Yekaterinburg 620137, Russia}
\affiliation{Department of Theoretical Physics and Applied Mathematics, Ural Federal University, 19 Mira St., Yekaterinburg 620002, Russia}
\author{Artem~R.~Oganov}
\affiliation{Skolkovo Institute of Science and Technology, 30 Bolshoy Boulevard, bld.~1, Moscow 121205, Russia}

\date{\today}

\begin{abstract}
On the basis of the first-principles evolutionary crystal structure prediction of stable compounds in the Cu–F system, we predict two experimentally unknown stable phases -- Cu$_2$F$_5$ and CuF$_3$. Cu$_2$F$_5$ comprises two interacting magnetic subsystems with the Cu atoms in the oxidation states +2 and +3. CuF$_3$ contains magnetic Cu$^{3+}$ ions forming a lattice with the antiferromagnetic coupling. We showed that some or all of Cu$^{3+}$ ions can be reduced to Cu$^{2+}$ by electron doping, as in the well known KCuF$_3$. Significant similarities between the electronic structures calculated in the framework of DFT+U suggest that doped CuF$_3$ and Cu$_2$F$_5$ may exhibit high-$T\mathrm{_c}$ superconductivity with the same mechanism as in cuprates.
\end{abstract}

\pacs{Valid PACS appear here}
\maketitle

Transition metal fluorides have been thoroughly studied during the last century~\cite{WINFIELD1986159, Thrasher}.  Among them coinage metal fluorides recently attracted considerable attention: Cu–F system in the electrochemistry field~\cite{Wang2011, Wang2012, Hua2014, Omenya2019}, Ag-F as a potentially new route to superconductivity~\cite{Grochala2001, Gawraczynski2019}, and Au-F due to the unusual oxidation state of gold~\cite{Lin2018}. Actually, Cu-F system contains an old puzzle of crystalline copper fluoride existence and synthesis, which produced a never-ending debate and remained unresolved to date. The first report of synthesis of CuF with zincblende structure was published in 1933~\cite{Ebert1933}. It was then argued that the reported CuF is identical to Cu$_2$O, dismissing the previous experimental results~\cite{Haendler1954}. Recently no one succeeded in reproducing the synthesis of CuF, and the earliest studies have met strong criticism~\cite{housecroft2005inorganic, greenwood2012chemistry}, since it is commonly believed that fluorine, because of its high electronegativity, will always oxidize copper to the oxidation state +2. Even though all attempts to synthesize CuF have been unsuccessful and the very existence of this compound is questionable, studies are ongoing~\cite{Wang2013_exp, Woidy2015}, and the complexes of CuF are already well characterized~\cite{Gulliver1981}. 

The computationally guided studies of new transition metal fluorides, and CuF in particular, also continue. Initially, they mainly compared different structure prototypes to find a hypothetical ground state crystal structure~\cite{Sohnel2005, Walsh2012} or investigated cluster formation~\cite{Krawczyk2006}. A variety of new structures have been reported using evolutionary crystal structure prediction and assuming CuF stoichiometry~\cite{Kuklin2019}. On the basis of all previous studies, eventually, it has been shown that all predicted structures are metastable~\cite{Walsh2012, Kuklin2019}.

Recently, a computational crystal structure prediction of coinage metal fluorides at different pressures was done~\cite{Lin2018}. However, the used method works with a fixed stoichiometry, which limits the prediction of new phases in the whole system. Moreover, redoing the same calculations lead to different structures~\cite{Liu2019}. Thus, the detailed and reliable analysis of the whole Cu-F system remained to be done.

In this Letter, we present a first-principles variable-composition evolutionary crystal structure prediction study of all phases in the Cu–F system. We recover the experimentally known structure of CuF$_2$ and report hitherto unknown stable $C2/m$-Cu$_2$F$_5$, $R\bar3c$-CuF$_3$, and $Pnma$-CuF$_3$ phases. Based on the similarities between the crystal structure of the discovered fluorides and the structure of the parent cuprate high-temperature (high-$T\mathrm{_c}$) superconductor La$_2$CuO$_4$, we explored the possibility of high-T$_s$ superconductivity in doped copper fluorides.

Stable phases in the Cu–F system were predicted here using the first-principles evolutionary algorithm as implemented in the USPEX package~\cite{Oganov2006, Oganov2011}. The evolutionary search was combined with structure relaxation and energy calculations using density functional theory (DFT) within the Perdew–Burke–Ernzerhof (PBE)~\cite{Perdew1996a} exchange–correlation functional and employing the projector augmented plane wave (PAW) method~\cite{Blochl1994} as implemented in the VASP package~\cite{VASP}. We used the plane-wave energy cutoff of 600~eV and $\Gamma$-centered $k$-meshes with a resolution of 2$\pi\times$0.05~Å$\textsuperscript{-1}$ for Brillouin zone sampling, ensuring excellent convergence of the quantities of interest. During the variable-composition structure search, the first generation of 160 structures was produced using random symmetric~\cite{Lyakhov2013} and random topological~\cite{Bushlanov2019} structure generators, with up to 18 atoms in the primitive cell. 70$\%$ of the next generation were obtained by applying variation operators (heredity, softmutation, lattice mutation) to the 70$\%$ of the lowest-energy structures of current generation and the other 30$\%$ of the generation were produced randomly. 

\begin{figure}[ht!]
\begin{minipage}[h]{1\linewidth}
\center{\includegraphics[trim={0cm 0.5cm 2cm 2cm},clip, width=1\linewidth]{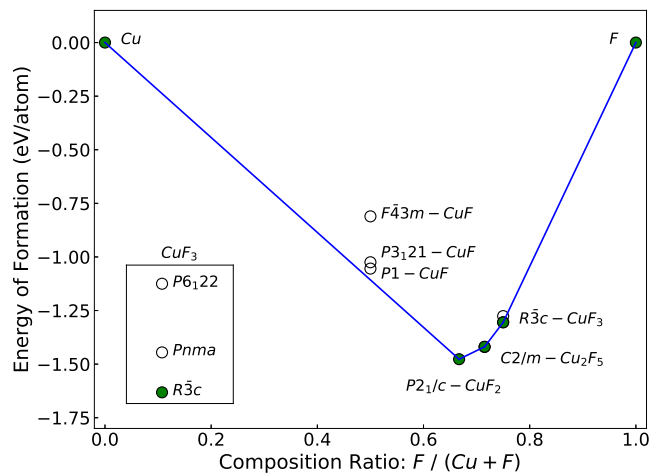}}
\end{minipage}
\caption {Convex hull diagram of the Cu–F system. The inset schematically mentions existence of three CuF$_3$ phases.}
\label{fig:cuf_ch}
\end{figure}

Phases located on the thermodynamic convex hull are stable with respect to decomposition into elemental Cu and F or other Cu–F compounds. The spin-polarized DFT calculations lead to the convex hull diagram as presented on the (Fig.~\ref{fig:cuf_ch}). It contains experimentally known $P2_1/c$-CuF$_2$, hitherto unknown $C2/m$-Cu$_2$F$_5$, $R\bar3c$-CuF$_3$, and slightly metastable $Pnma$-CuF$_3$, which is just 0.001~eV above the convex hull. Successful prediction of the CuF$_2$, a known compound, indicates the robustness of our methodology. All obtained potentially stable structures became subject of an additional fixed-composition study taking into account up to four formula units (and up to 18 atoms in the unit cell for CuF). The dynamical stability of all structures was carefully verified with phonon calculations using the supercell approach and the finite displacement method, as implemented in the Phonopy package~\cite{Togo2015}. The structural information for the obtained compounds and results of the phonon calculations are presented in the Supporting Materials (SM).

The most energetically favorable structure of CuF, found in our study, is the low-symmetry $P1$-CuF, which is even lower in energy than previous reports~\cite{Kuklin2019} by $\sim$0.05~meV/atom, but its low symmetry and high energy ($\sim$50~meV/atom above the convex hull) indicated its instability and tendency to decompose into Cu+CuF$_2$. Thus, we conclude CuF is unlikely to exist at ambient pressure. 

\begin{figure}[h!]
\begin{minipage}[h]{0.9\linewidth}
\center{\includegraphics[width=1\linewidth]{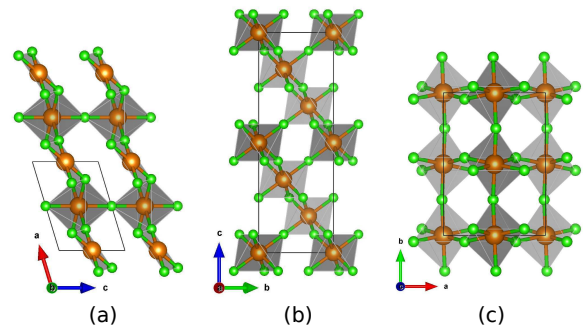}}
\end{minipage}
\caption {(color online). Schematic representation of the crystal structures: (a)~$C2/m$-Cu$_2$F$_5$, (b)~$R\bar3c$-CuF$_3$, (c)~$Pnma$-CuF$_3$. The Cu and F atoms are shown in brown and green, respectively. Structures were visualized using VESTA software~\cite{Momma2008}.}
\label{fig:crystal_structures}
\end{figure}

Cu$_2$F$_5$ crystallizes in the monoclinic space group $C2/m$ with two inequivalent Cu sites, where each Cu atom of the first type is bonded to six pairwise equivalent F atoms forming a CuF$_6$ octahedron~(Fig.~\ref{fig:crystal_structures}a), with the corner-sharing octahedral tilt angles of 0$^{\circ}$. In the second site, the Cu atom is in a square planar geometry with four pairwise equivalent F atoms. This arrangement could be also described as a distorted octahedron (see SM Fig.~3a). While isostoichiometric $P\bar1$-Ag$_2$F$_5$ is well-known~\cite{Grochala2001}, hypothetical $P\bar1$-Cu$_2$F$_5$ has a higher energy than $C2/m$-Cu$_2$F$_5$ by $\sim$3~meV/atom in the spin-polarized DFT solution.

Ground state CuF$_3$ has a trigonal perovskite structure with the space group $R\bar3c$ (Fig.~\ref{fig:crystal_structures}b). This structure was also predicted in~\cite{Lin2018}. The Cu atom is bonded with six equivalent F atoms to form an octahedron with the corner-sharing octahedral tilt angles of 29$^{\circ}$. Orthorhombic $Pnma$-CuF$_3$, metastable at 0~K, also has a perovskite structure ABX$_3$ with absent A cations -- ReO$_3$-type structure (Fig.~\ref{fig:crystal_structures}c), with the corner-sharing octahedral tilt angles of 28$^{\circ}$. Metal trifluorides FeF$_3$, CoF$_3$, RuF$_3$, RhF$_3$ PdF$_3$, and IrF$_3$ also have perovskite structure with the space group $R\bar3c$~\cite{Hepworth1957}, whereas AgF$_3$ and AuF$_3$ crystallize in a totally different structure with the space group $P6_122$~\cite{Zemva1991, Einstein1967}. Hypothetical $P6_122$-CuF$_3$ has a higher energy than the $R\bar3c$ phase by $\sim$30~meV/atom in the spin-polarized DFT solution. Notably, perovskite-type structures frequently have octahedral tilt instabilities and exhibit phase transition ~\cite{Woodward1997}.

Discovered Cu fluorides have significant crystal-chemical similarities with high-T$_s$ cuprates. In both systems we observe Cu$^{+2}$ (in square planar coordination, as a consequence of Jahn-Teller distortion in all cuprates and Cu$_2$F$_5$) and Cu$^{+3}$ (in CuF$_3$, Cu$_2$F$_5$, and in doped cuprates). As we discuss below, pure parent compounds CuF$_3$, Cu$_2$F$_5$, and La$_2$CuO$_4$ are antiferromagnetic insulators, but doping by electrons or holes makes them metallic and superconducting (for sure La$_2$CuO$_4$, and most likely for copper fluorides).

\begin{figure}[ht!]
\begin{minipage}[h]{0.49\linewidth}
\center{\includegraphics[width=1\linewidth]{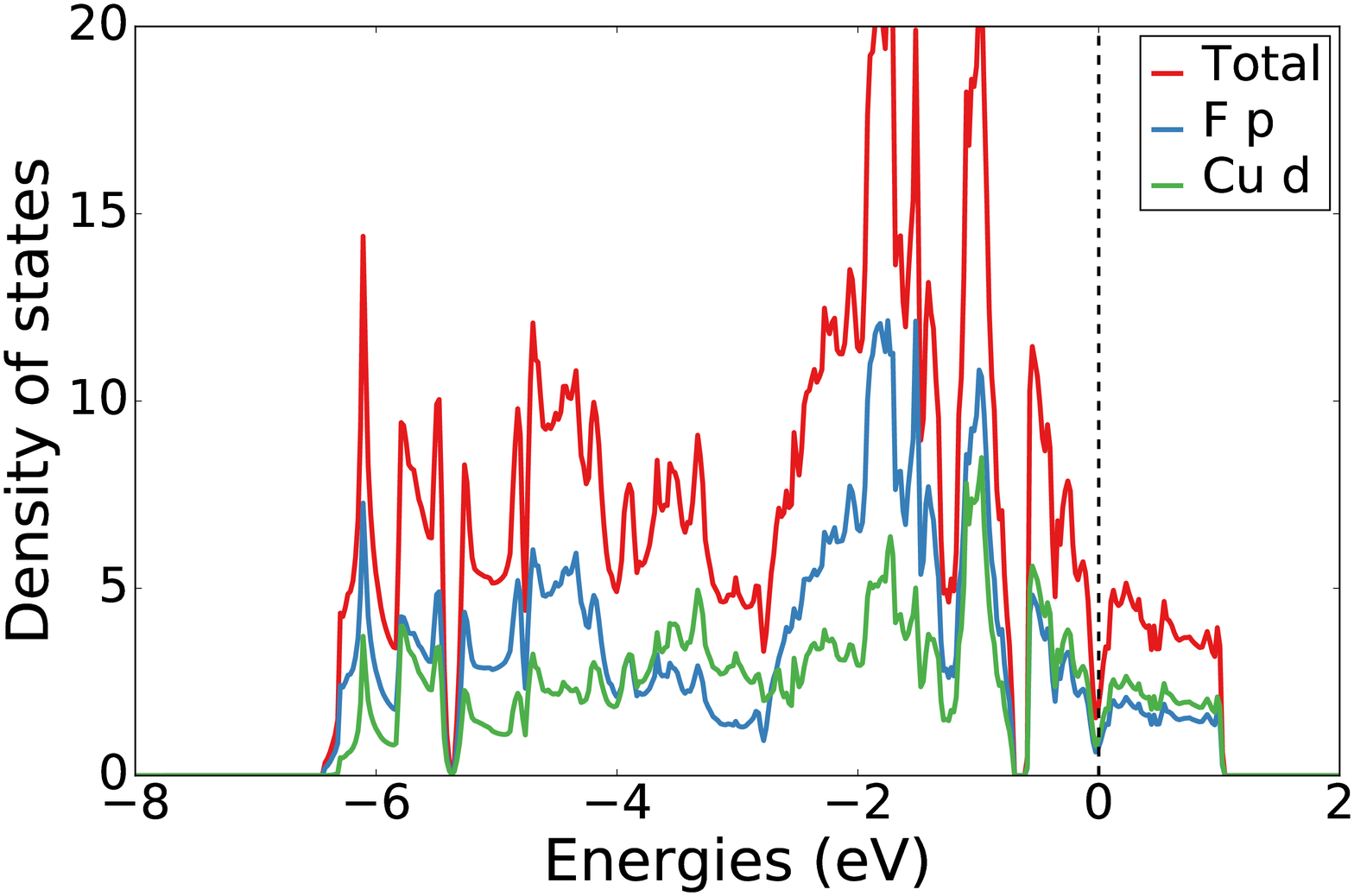}} (a) \\
\end{minipage}
\hfill
\begin{minipage}[h]{0.49\linewidth}
\center{\includegraphics[width=1\linewidth]{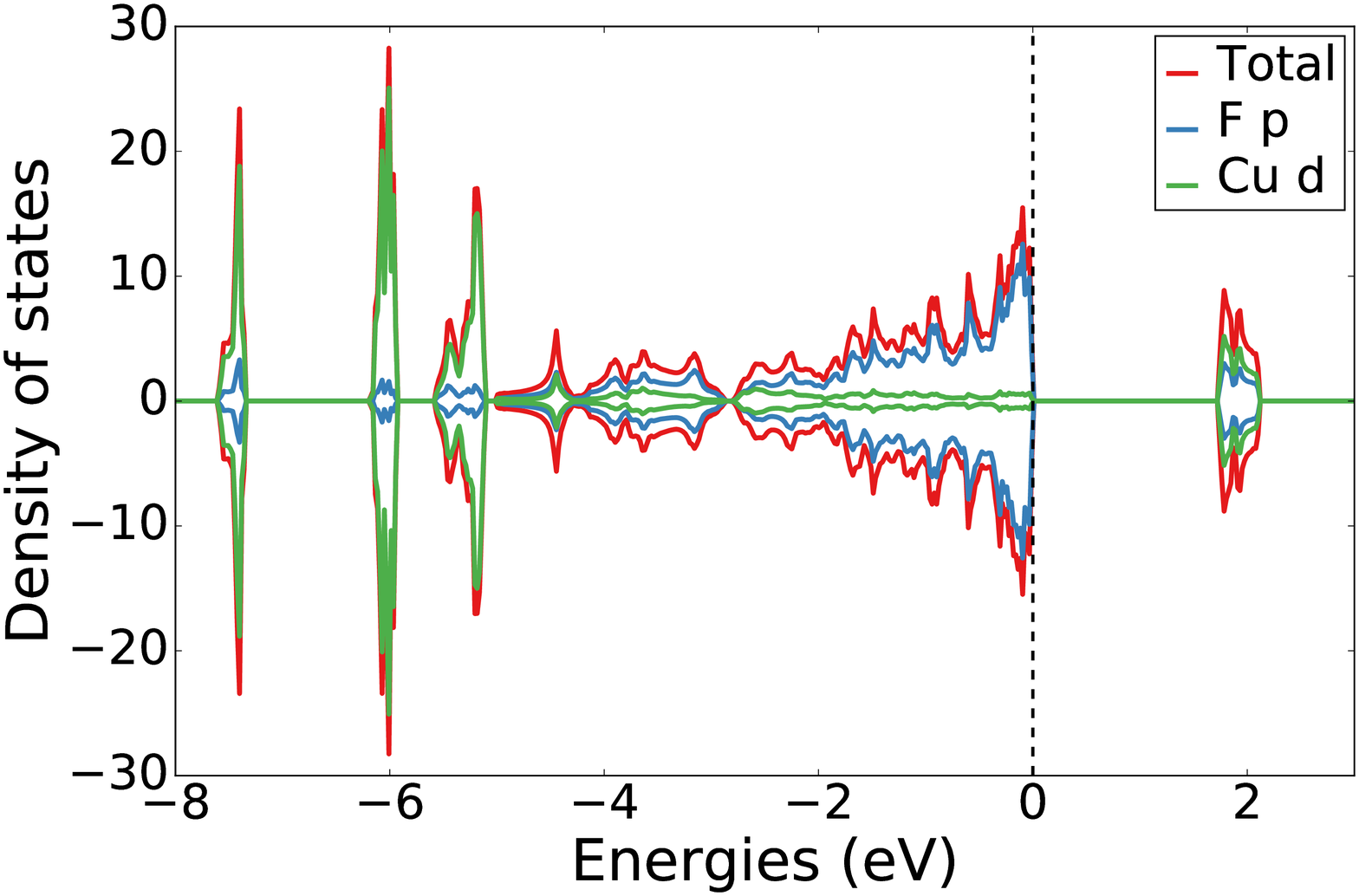}} (b) \\
\end{minipage}
\begin{minipage}[h]{0.49\linewidth}
\center{\includegraphics[width=1\linewidth]{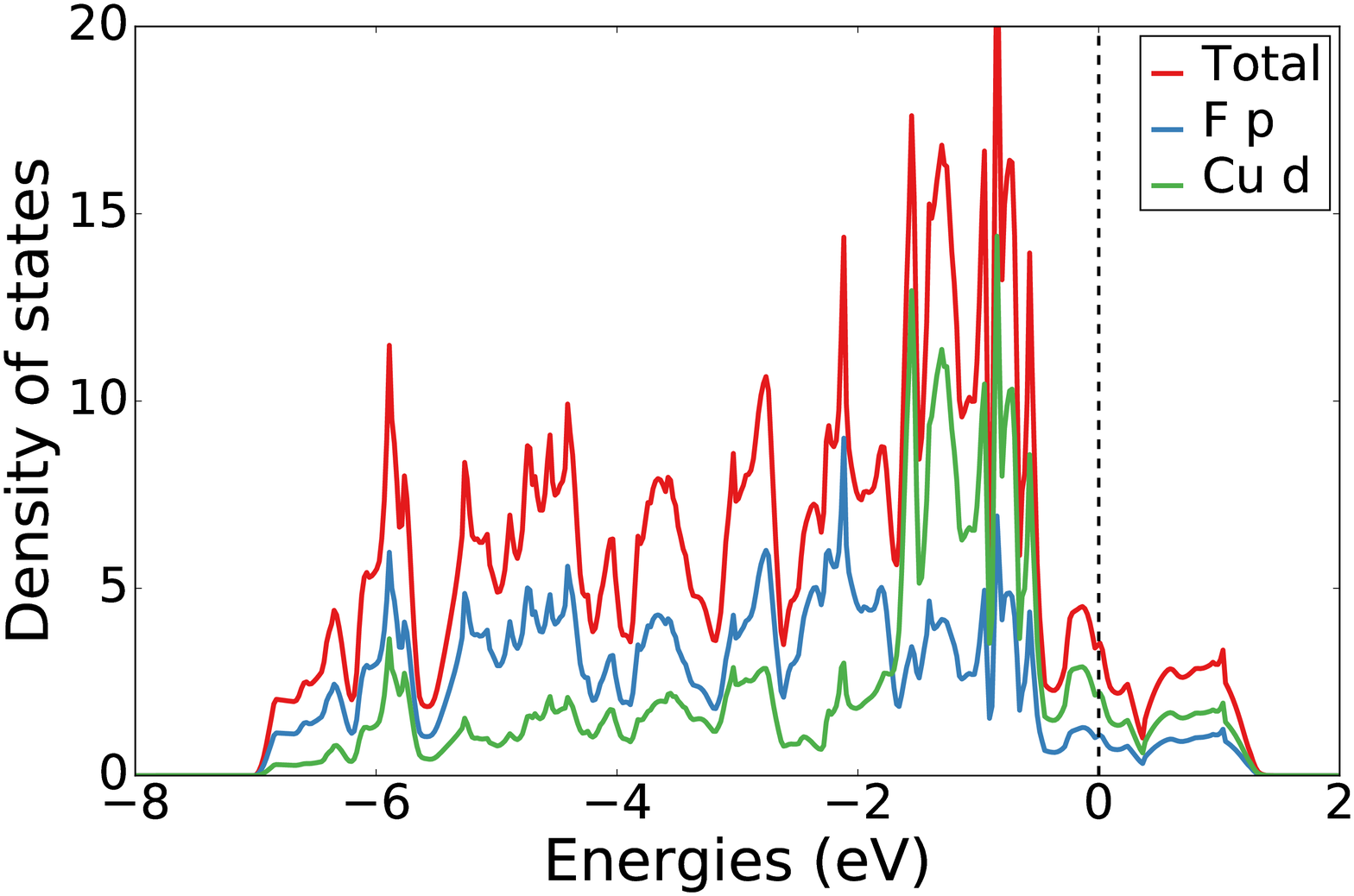}} (c) \\
\end{minipage}
\hfill
\begin{minipage}[h]{0.49\linewidth}
\center{\includegraphics[width=1\linewidth]{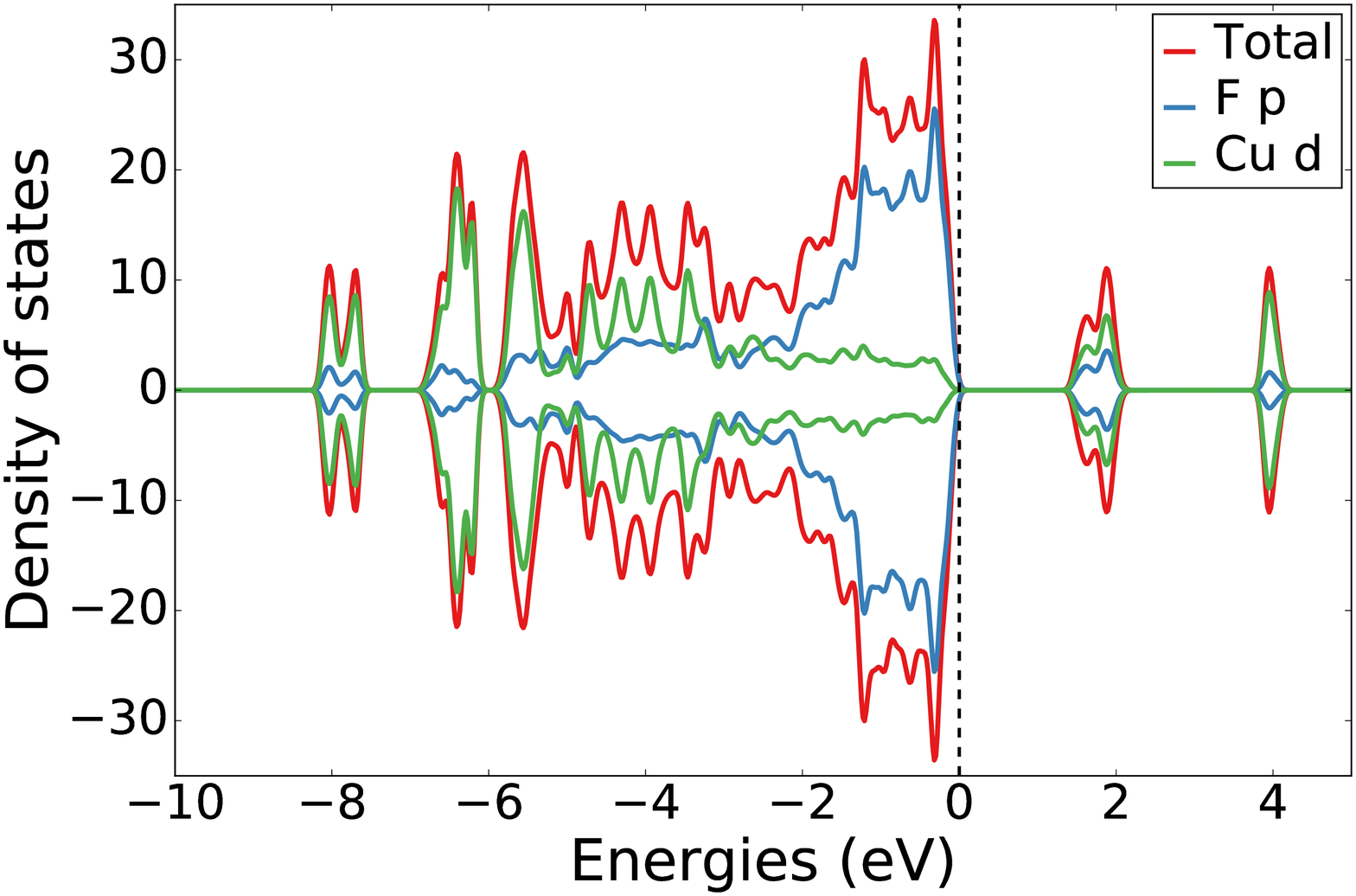}} (d) \\
\end{minipage}
\begin{minipage}[h]{0.49\linewidth}
\center{\includegraphics[width=1\linewidth]{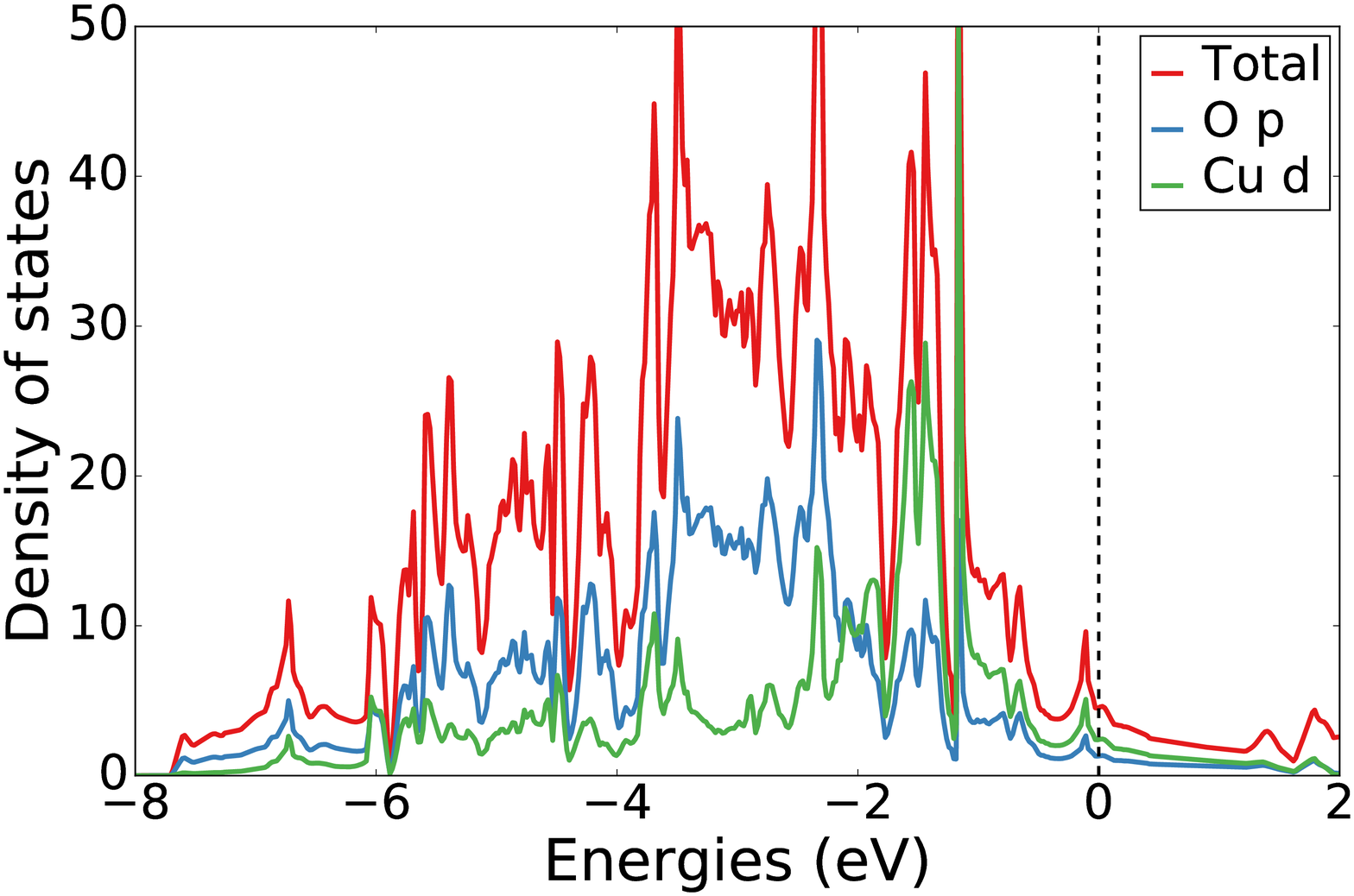}} (e) \\
\end{minipage}
\hfill
\begin{minipage}[h]{0.49\linewidth}
\center{\includegraphics[width=1\linewidth]{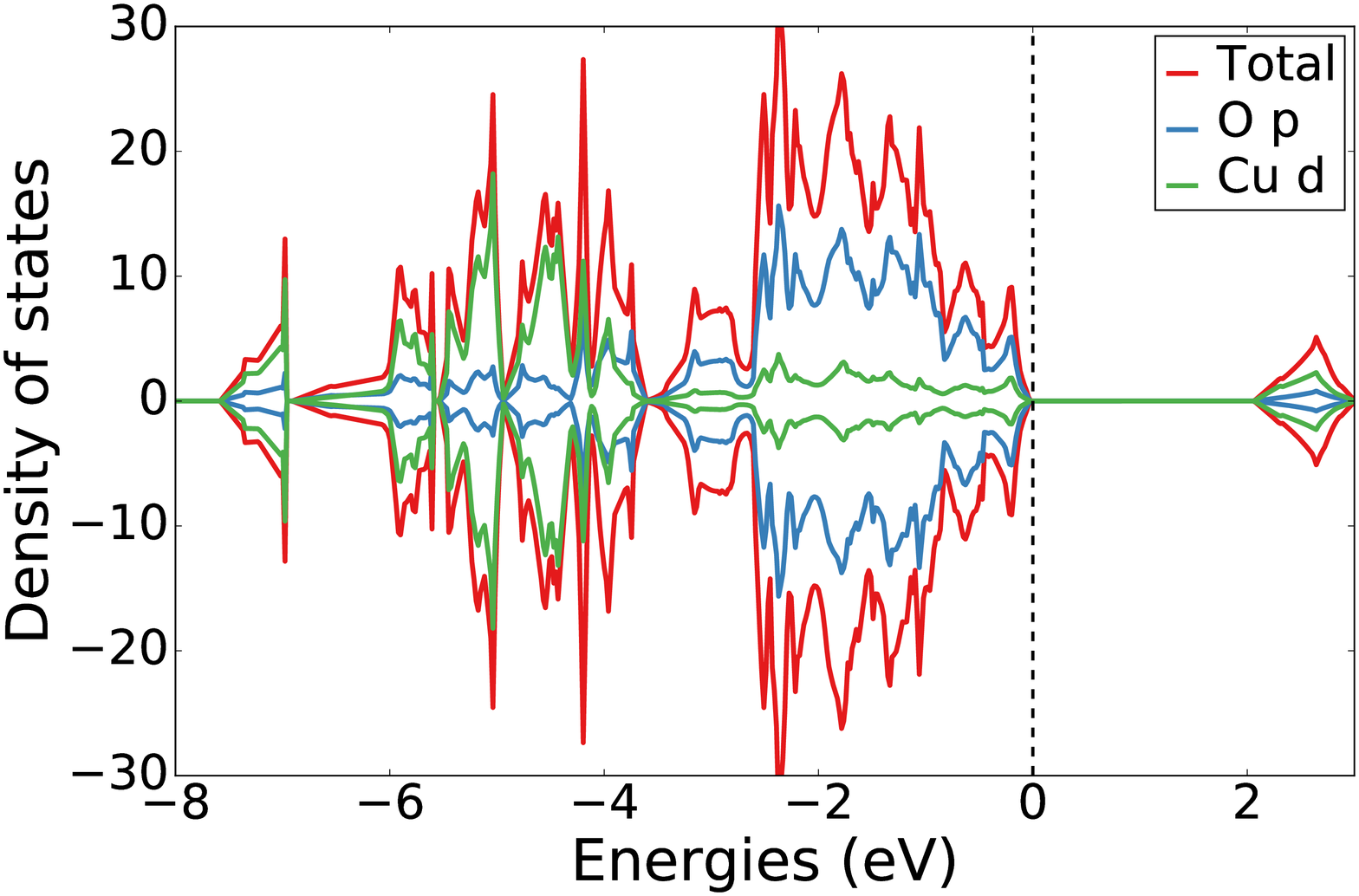}} (f) \\
\end{minipage}
\caption {(color online). Total and partial density of states for (a,~b) $R\bar3c$-CuF$_3$, (c,~d) Cu$_2$F$_5$, and (e,~f) $Bamb$-La$_2$CuO$_4$ obtained using (a,~c,~e) DFT and (b,~d,~f) DFT+U.} 
\label{fig:pdos_of_CuF3_r3c_and_La2CuO4}
\end{figure}

To compare the electronic properties of copper fluorides and cuprate, we firstly performed the spin-unpolarized DFT calculations using dense Monkhorst–Pack meshes of 8$\times$8$\times$4 and 12$\times$12$\times$12 kpoints for distorted orthorombic low-temperature $Bamb$-La$_2$CuO$_4$ phase and all fluorides, respectively. Structural information and energies for ferromagnetic and antiferromagnetic orders are presented in the (SM~Tab.1,~2). The densities of states (DOS) for $R\bar3c$-CuF$_3$, Cu$_2$F$_5$, and $Bamb$-La$_2$CuO$_4$ resolved for the Cu-d and ligand-p states within DFT are shown on (Fig.~\ref{fig:pdos_of_CuF3_r3c_and_La2CuO4}a,~c,~e). In all systems DFT results show that the Cu-3d energy band is located completely inside the p band of the ligands and strongly hybridizes with it. Therefore, the partially filled electronic states of interest are formed by the d- and p-symmetry states with approximately equal weights, and the usual ionic picture is not applicable for such a band structure. Notably, the magnetic exchange interaction is proportion to the scale of magnetic fluctuations. For all considered systems, solutions with antiferromagnetic order are the lowest in energy (SM Tab.3,~4). Since the energy difference obtained from the spin-unpolarized and spin-polarized DFT calculations is quite small, one can expect strong spin fluctuations in both types of systems -- and we recall that high-T$_c$ superconductivity of cuprates is believed to be mediated by spin fluctuations. Doped Cu fluorides can, or perhaps, even should be superconducting by the same magnetically mediated mechanism.

Although DFT shade light on some premature analogy with cuprates, in principle, this method is pathological since cannot correctly reproduce the antiferromagnetic insulating state of La$_2$CuO$_4$ because it neglects on-site Coulomb correlations~\cite{Czyyk1994}, and more robust results are achieved by taking into account the electronic correlations using the DFT+U method with the Coulomb interaction parameter $U=8$~eV and the exchange interaction parameter $J=0.9$~eV~\cite{Czyyk1994, Anisimov1997}. Because we deal with copper in the same divalent and trivalent states, and the energy bands in cuprates and copper fluorides studied here have similar widths (Fig.~\ref{fig:pdos_of_CuF3_r3c_and_La2CuO4}a,~c,~e and Tab.~\ref{parameters}), we chose the same values of $U$ and $J$ for all calculations taking into account the on-site Coulomb repulsion between the Cu-3d electrons in CuF$_3$, Cu$_2$F$_5$, and $Bamb$-La$_2$CuO$_4$. Structural information obtained after the relaxation with DFT+U as well as the values of total energy for ferromagnetic and antiferromagnetic orders are presented in the (SM~Tab.1,~2).

We reproduced the insulating antiferromagnetic ground state of $Bamb$-La$_2$CuO$_4$ with the DFT+U energy gap of about 2~eV and the magnetic moment of the Cu atoms of $0.65~\mu_\mathrm{B}$, which is in a close agreement with the experimentally observed values of $\sim$2~eV and $0.68~\mu_\mathrm{B}$, respectively~\cite{Pickett1989}. The DOS for $R\bar3c$-CuF$_3$, Cu$_2$F$_5$, and $Bamb$-La$_2$CuO$_4$ obtained using DFT+U are presented on (Fig.~\ref{fig:pdos_of_CuF3_r3c_and_La2CuO4}b,~d,~f). The DOS for $Pnma$ phase is presented on the (Fig.~2a,~b in SM). For $R\bar3c$-CuF$_3$ and $Pnma$-CuF$_3$ in the antiferromagnetic phase, the DFT+U calculations show similarities in the key features of the electronic structures of CuF$_3$ and $Bamb$-La$_2$CuO$_4$ -- they have well-separated Hubbard bands formed by the Cu-d states, whereas the first ionization states have a p-symmetry and are formed by the ligands. 

In the CuF$_3$ structures, copper has an atypical formal oxidation state of +3, which leads to the 3d$^8$ electronic configuration, whereas in $Bamb$-La$_2$CuO$_4$ there are Cu$^{2+}$ Jahn–Teller active ions. However, some or all of Cu$^{+3}$ ions in CuF$_3$ can be reduced to Cu$^{+2}$ by electron doping, like in the well-known perovskite-type KCuF$_3$, where all Cu atoms are in the oxidation state +2. CuF$_3$, in fact, can be described as the structure of KCuF$_3$ with all K atoms removed. Thus, one way to make a superconducting Cu fluoride is to remove part of K atoms from KCuF$_3$ (in a vacuum tube) -- the result should be a metallic perovskite-type compound with mixed Cu$^{+2}$ and Cu$^{+3}$ states. To clearly show this, we performed a fixed-composition structure search of K$_3$(CuF$_3$)$_4$, which determined that the most stable phase has perovskite-type structure with the space group $Im\bar3m$ (SM Fig.~3c). This structure is stable with respect to the decomposition into $R\bar3c$-CuF$_3$ and KCuF$_3$ ($\sim$0.05~eV/atom below the decomposition line), which means that potassium ions can be easily extracted from the KCuF$_3$, forming mixed-valence compound.

\begin{table}[hb!]
\caption{Bandwidth $W$ and charge transfer gap $\Delta_\mathrm{pd}$ calculated using DFT. Hubbard bands splitting $U_\mathrm{dd}$, spin $S$, and magnetic moment $M$ obtained using the DFT+U method. The values in parentheses are related to the second type of Cu atoms in the Cu$_2$F$_5$.}
\begin{tabular}{l | c c c c c}
\hline
\hline
  & $W$ (eV) &  $\Delta_\mathrm{pd}$ (eV)  & $U_\mathrm{dd}$ (eV) & $S$ & $M$ ($\mu_\mathrm{B}$)  \\
\hline
 $R\bar3c$-CuF$_3$ & 8 & 1.7 & 9.5 & 1 & 1.15 \\
 Cu$_2$F$_5$ & 8 & 1.42 & 9.5 & 1 (1/2) & 1.17 (0.79)\\
 La$_2$CuO$_4$ & 9 & 2 & 10.5 & 1/2 & 0.65\\
\hline
\hline
\end{tabular}
\label{parameters}
\end{table}

\begin{figure}[ht!]
\begin{minipage}[h]{0.49\linewidth}
\center{\includegraphics[width=1\linewidth]{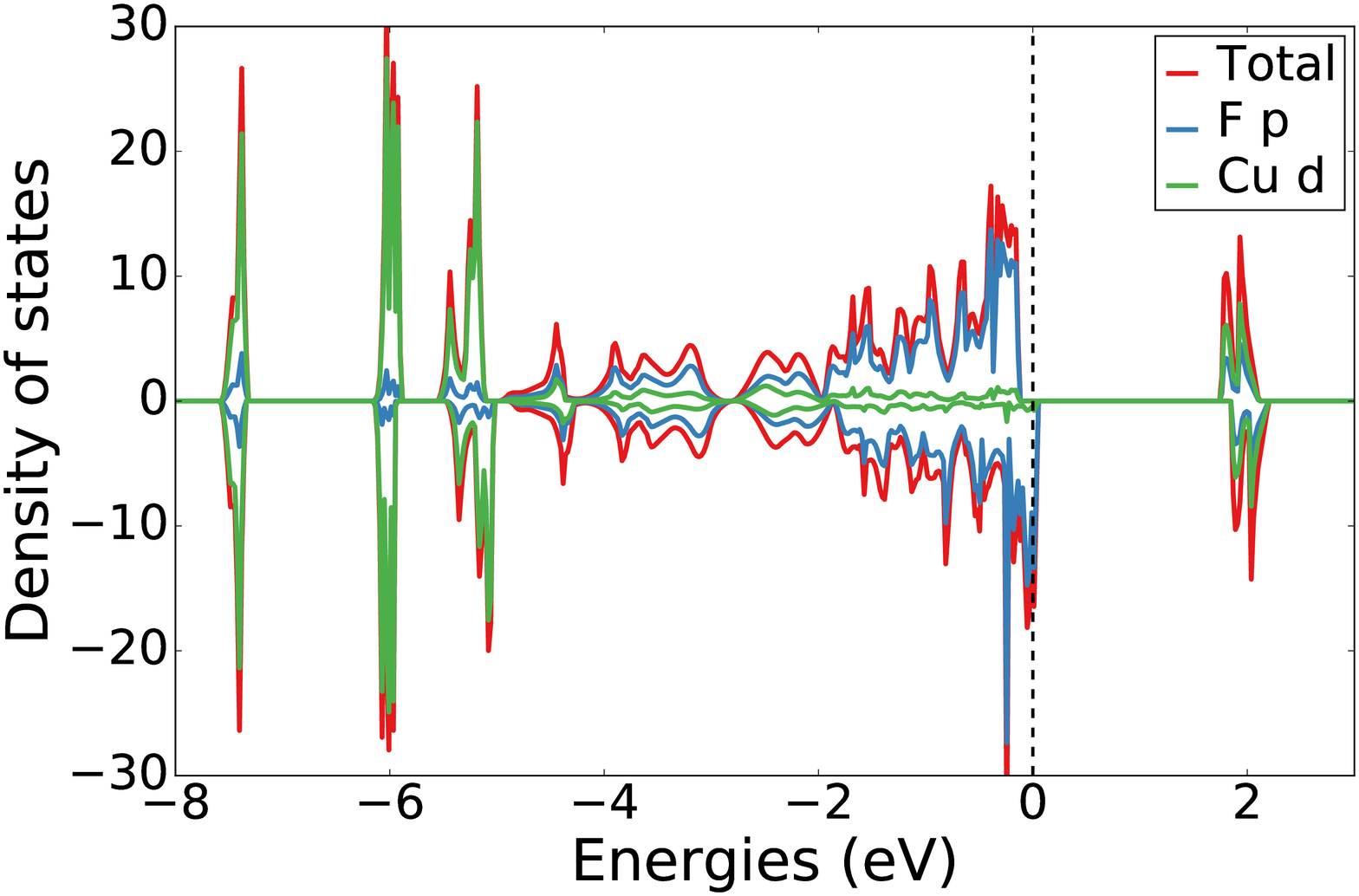}} (a) \\
\end{minipage}
\hfill
\begin{minipage}[h]{0.49\linewidth}
\center{\includegraphics[width=1\linewidth]{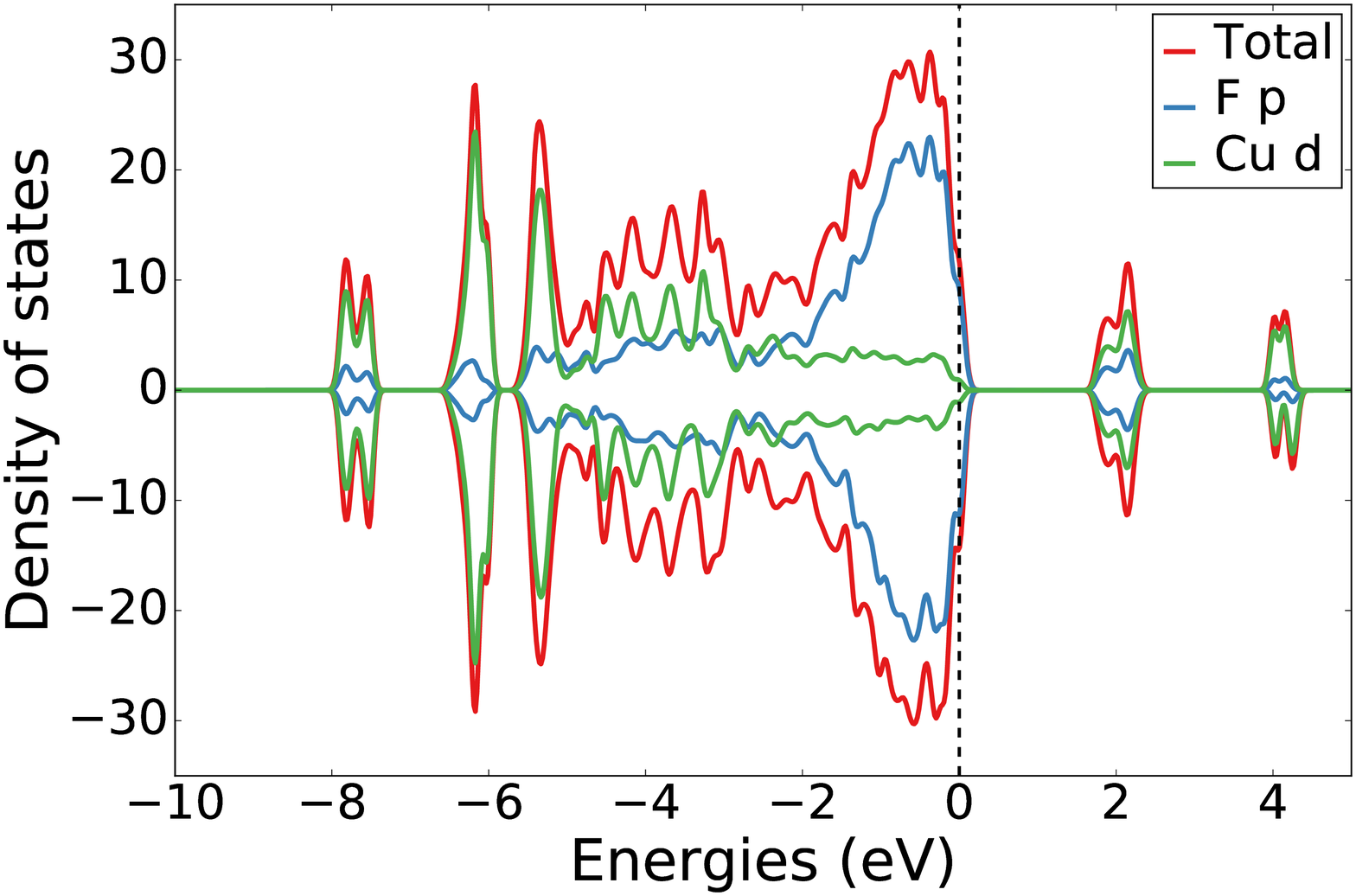}} (b) \\
\end{minipage}
\begin{minipage}[h]{0.49\linewidth}
\center{\includegraphics[width=1\linewidth]{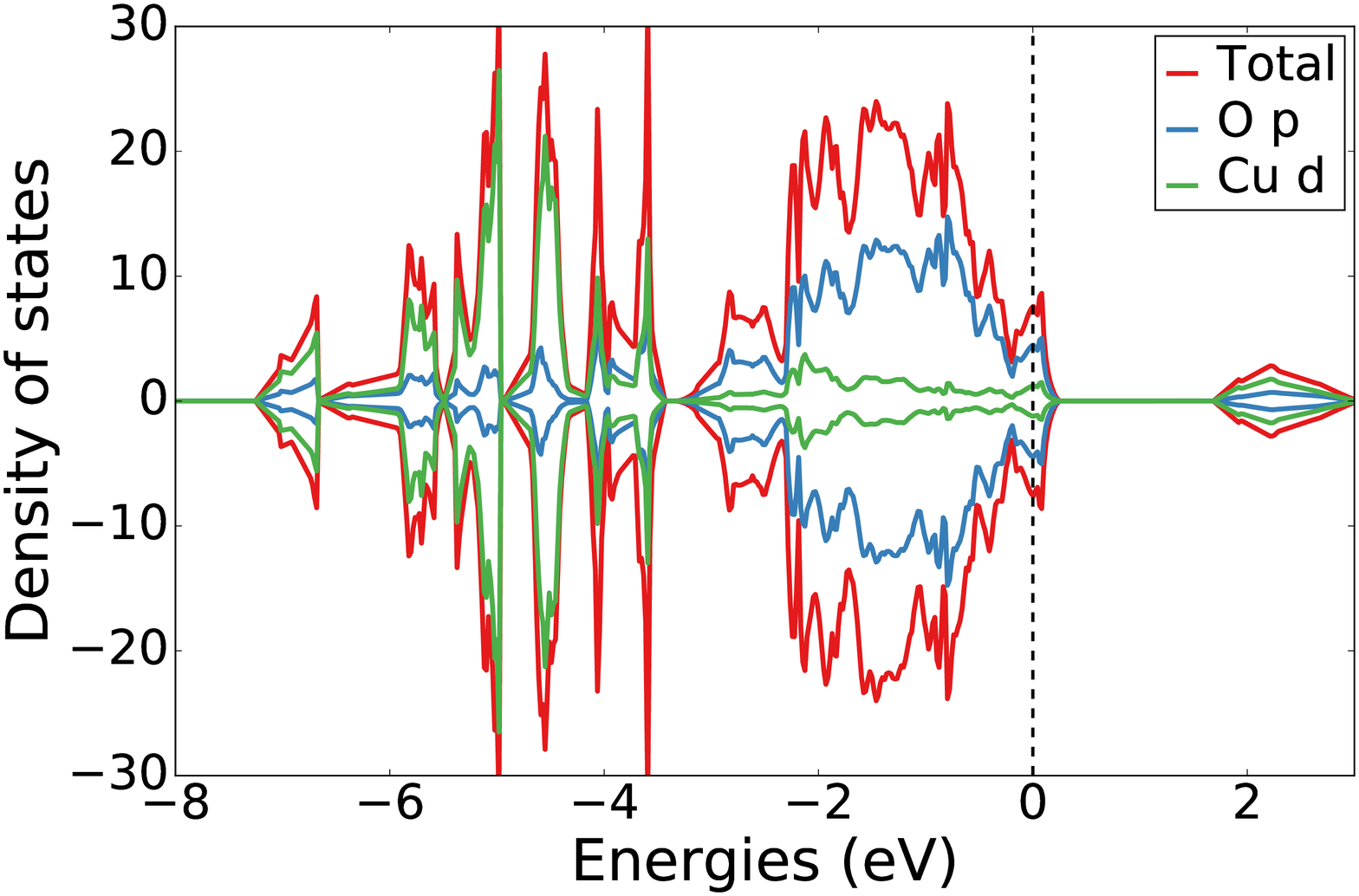}} (c) \\
\end{minipage}
\hfill
\begin{minipage}[h]{0.49\linewidth}
\center{\includegraphics[width=1\linewidth]{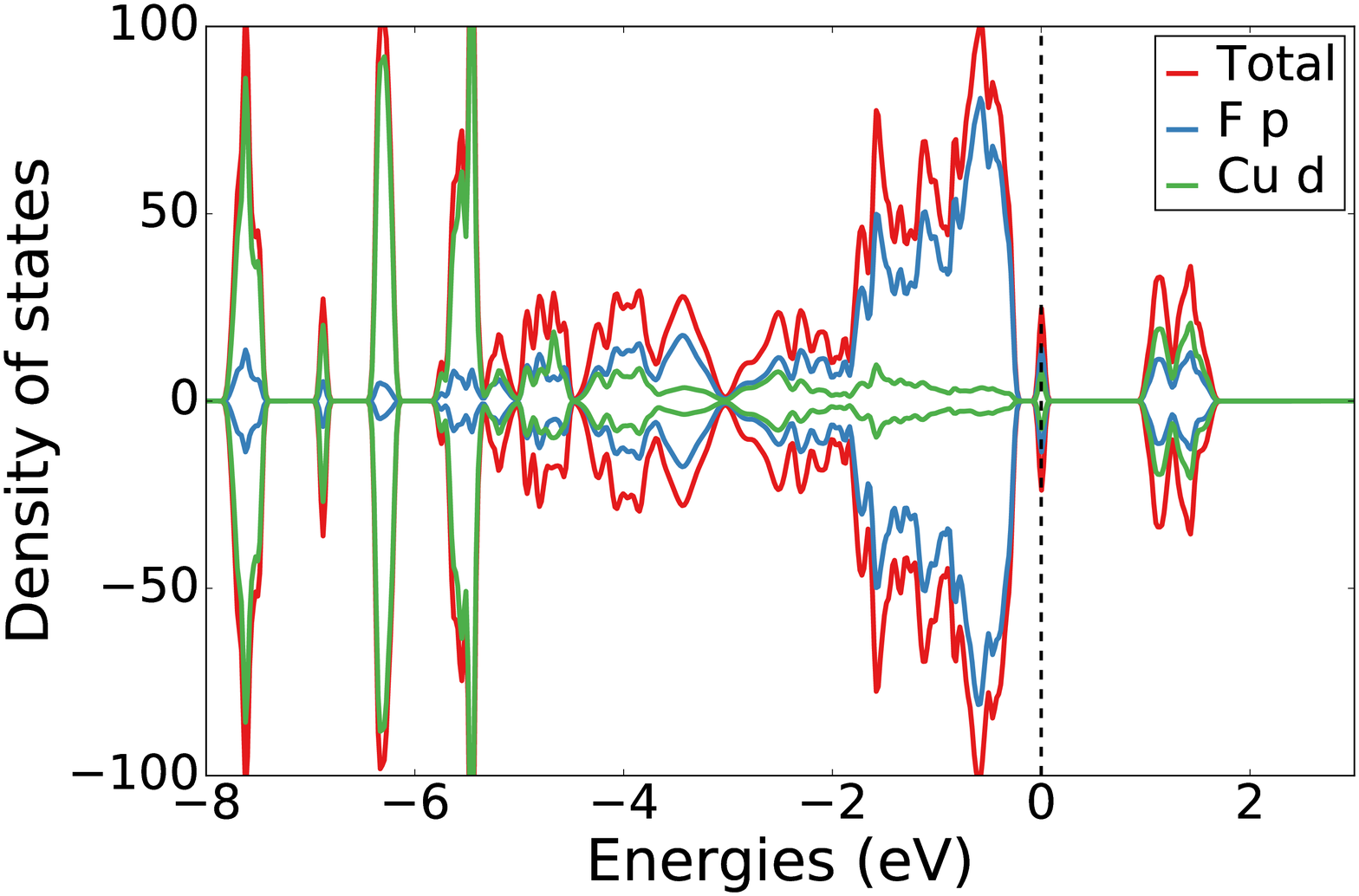}} (d) \\
\end{minipage}
\begin{minipage}[h]{0.49\linewidth}
\center{\includegraphics[width=1\linewidth]{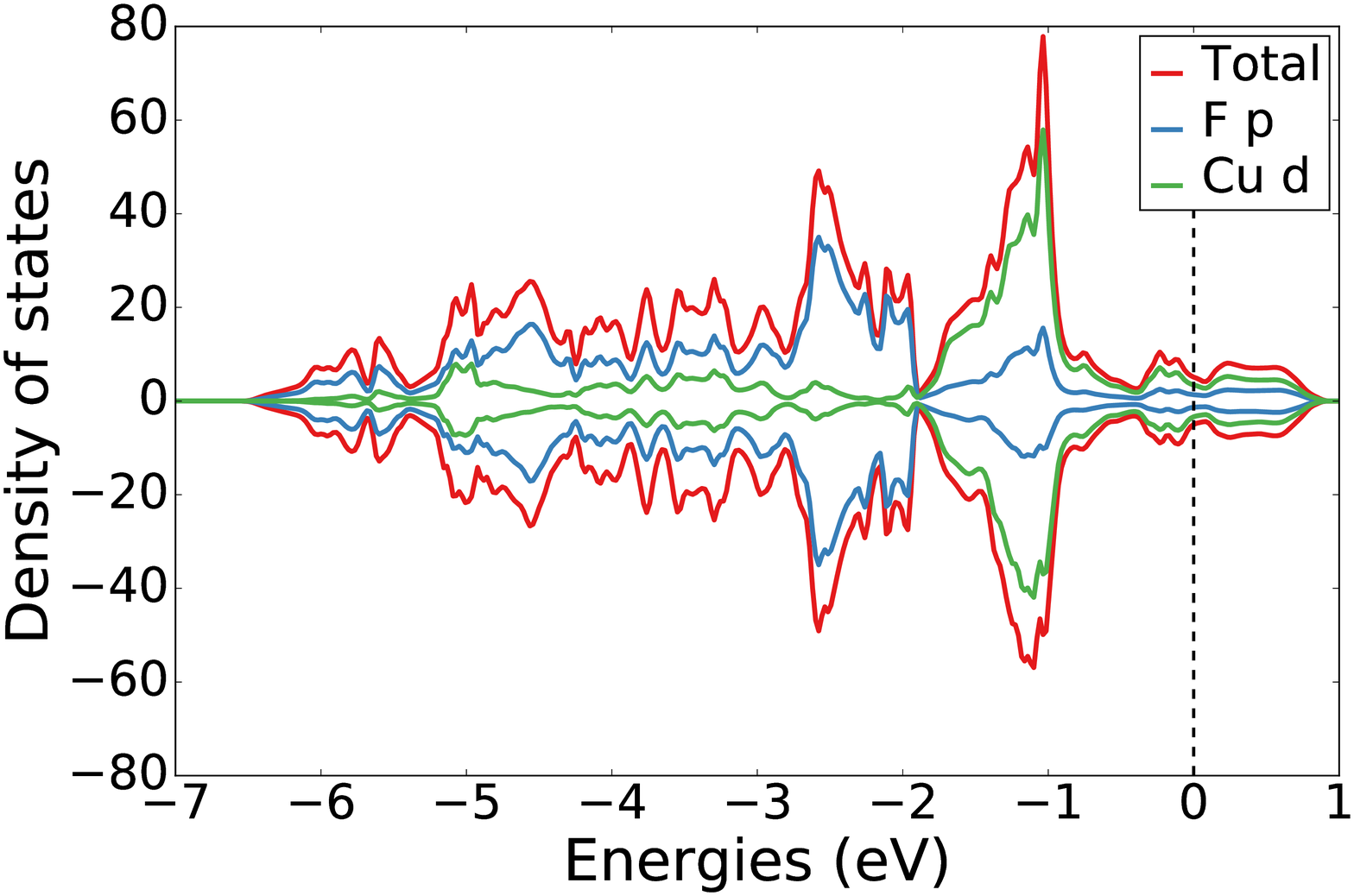}} (e) \\
\end{minipage}
\caption{(color online). Total and partial density of states for p-doped (a)~$R\bar3c$-CuF$_3$, (b)~Cu$_2$F$_5$, and (c)~La$_2$CuO$_4$. (d)~$R\bar3c$-CuF$_3$ with the one F vacancy per 2$\times$2$\times$2 supercell, (e)~and perovskite-type Im$\bar3$m-K$_3$(CuF$_3$)$_4$.}
\label{fig:hole_doping_DOS}
\end{figure}

According to the DFT+U solution: each Cu site in the CuF$_3$ has spin 1; Cu$_2$F$_5$ is determined as a compound with the mixed-valence Cu$^{2+}$/Cu$^{3+}$ state and Cu ions with spin 1 and 1/2; Cu ions in the cuprate have spin 1/2. The magnetic moments per Cu atom obtained in the DFT+U calculations for $R\bar3c$-CuF$_3$ are 1.15~$\mu_\mathrm{B}$ (1.14~$\mu_\mathrm{B}$ in the $Pnma$ phase). These magnetic moment values are smaller by a factor of 0.58 than the formal ionic value of 2~$\mu_\mathrm{B}$ for a Cu$^{3+}$ ion compared to the reduction factor of 0.65 for the formal atomic value of 1~$\mu_\mathrm{B}$ for a Cu$^{2+}$ ion in the La$_2$CuO$_4$~\cite{Pickett1989}. For Cu$_2$F$_5$ we found that two types of Cu atoms, have different formal electronic configurations, d$^8$ and d$^9$, and different magnetic moments of 1.17~$\mu_\mathrm{B}$ and 0.79~$\mu_\mathrm{B}$. All predicted copper fluoride structures and $Bamb$-La$_2$CuO$_4$ are charge-transfer insulators with respect to the classification of Zaanen \textit{et~al.}~\cite{Zaanen1985}. The first energy excitation occurs between the p band of the ligands and the d band of the metal ion. CuF$_3$ has a small charge transfer gap (an important characteristic of cuprates) $\Delta_\mathrm{pd}$=1.7~eV, Cu$_2$F$_5$ also has a small charge transfer gap $\Delta_\mathrm{pd}$=1.42~eV, comparable with 2~eV of La$_2$CuO$_4$. Though the energy gap depends on the choice of the Hubbard $U$ parameter, the charge-transfer nature of the gap remains the same for a wide range of $U$ values in fluorides. The splitting between the Hubbard bands for CuF$_3$ and La$_2$CuO$_4$ is similar and equals to $\sim$9.5~eV and 10.5~eV, respectively (Fig.~\ref{fig:pdos_of_CuF3_r3c_and_La2CuO4}b,~d).

Cu$_2$F$_5$, CuF$_3$, and La$_2$CuO$_4$ are insulators and can display superconductivity only when properly doped. Consequently, we performed DFT+U calculations of the considered systems doped with holes using rigid-band shift approximation, with doping amounted to 0.25 holes for each copper atom in the unit cell (Fig.~\ref{fig:hole_doping_DOS}a,~b,~c). CuF$_3$ and Cu$_2$F$_5$, like La$_2$CuO$_4$, undergo a transition from the insulating to the conducting state upon the hole doping, which once again highlights the similarity of their electronic properties. We have also examined a 2$\times$2$\times$2 $R\bar3c$-CuF$3$ supercell (64~atoms) with a vacancy on one of the F atoms. This ferrimagnetic structure lies on the thermodynamic convex hull (Fig.~\ref{fig:cuf_ch}). This means that the formation of the non-stoichiometric CuF$_{3-x}$ is favorable. The DFT+U solution determines the formation of a peak at the Fermi level for this structure (Fig.~\ref{fig:hole_doping_DOS}~d).  Perovskite-type $Im\bar3m$-K$_3$(CuF$_3$)$_4$ also has a metallic solution in DFT+U study (Fig.~\ref{fig:hole_doping_DOS}~e).

In summary, the results of the systematic crystal structure search in the Cu–F system supports that CuF is unlikely to exist and have revealed hitherto unknown $C2/m$-Cu$_2$F$_5$, $R\bar3c$-CuF$_3$, and slightly metastable $Pnma$-CuF$_3$. Cu$_2$F$_5$ contains Cu ions with oxidation states +2 and +3, which leads to the presence of two magnetic subsystems. In CuF$_3$, Cu ions have an unusual oxidation state +3, which can be reduced to +2 by proper doping. We showed that potassium can be extracted from KCuF$_3$ forming metallic state. we showed using DFT+U that all discovered copper fluorides are strongly correlated compounds and charge-transfer insulators. Since comparison of the CuF$_3$ and Cu$_2$F$_5$ with the classical cuprate La$_2$CuO$_4$ shows many similarities, discovered structures possibly could be a new class of high-$T\mathrm{_c}$ superconductors.

{\it Acknowledgments.}
This work was supported by the Russian Science Foundation (Project 19-72-30043). Calculations were performed on Arkuda cluster of Skoltech and Uran cluster of IMM UB RAS. DYN, DMK and VIA thanks the Ministry of Science and Higher Education of the Russian Federation (No.~AAAA-A18-118020190098-5, topic "Electron").

\bibliography{main}
\end{document}


\date{}
\maketitle 

\begin{table}[h!]
\caption{Structural parameters of $C2/m$-Cu$_2$F$_5$ optimized using DFT (left) and DFT+U (right).}
\parbox{.45\linewidth}{
\centering
\begin{tabular}{c c c c c}
\hline
\hline
Lattice & Atoms & $x$ & $y$ & $z$ \\
\hline
 a = 5.16          &  Cu1  &  0.50 & 0.50 & 0.50 \\
 b = 5.16          &  Cu2  &  0.00 & 0.00 & 0.50 \\
 c = 3.78          &  F1   &  0.19 & 0.19 & 0.27 \\
 $\alpha$ = 104.80   &  F2   &  0.80 & 0.80 & 0.72 \\
 $\beta$  = 104.80   &  F3   &  0.23 & 0.76 & 0.50 \\
 $\gamma$ = 95.10    &  F4   &  0.76 & 0.23 & 0.50 \\
 V = 93.1          &  F5   &  0.50 & 0.50 & 0.00 \\
\hline
\hline
\end{tabular}
}
\hfill
\parbox{.45\linewidth}{
\centering
\begin{tabular}{c c c c c}
\hline
\hline
Lattice & Atoms & $x$ & $y$ & $z$ \\
\hline
 a = 5.0165          &  Cu1  &  0.50 & 0.50 & 0.50 \\
 b = 5.0165          &  Cu2  &  0.00 & 0.00 & 0.50 \\
 c = 3.6715          &  F1   &  0.80  &  0.80  &  0.75 \\
 $\alpha$ = 106.44   &  F2   &  0.19  &  0.19  &  0.24 \\
 $\beta$  = 106.44   &  F3   &  0.24  &  0.75  &  0.50 \\
 $\gamma$ = 95.07    &  F4   &  0.75  &  0.24  &  0.50 \\
 V = 83.62~\AA$^{3}$ &  F5   &  0.50  &  0.50  &  0.00 \\
\hline
\hline
\end{tabular}
}
\end{table}

\begin{table}[h!]
\caption{Structural parameters of $R\bar3c$-CuF$_3$ optimized using DFT (left) and DFT+U (right).}
\parbox{.45\linewidth}{
\centering
\begin{tabular}{c c c c c}
\hline
\hline
Lattice & Atoms & $x$ & $y$ & $z$ \\
\hline
a = 5.30          &  Cu1  &  0.50 &   0.50 &  0.50 \\
b = 5.30          &  Cu2  &  0.00 &   0.00 &  0.00 \\
c = 5.30          &  F1   &  0.65 &   0.84 &  0.25 \\
$\alpha$ = 58.2     &  F2 &  0.34 &   0.15 &  0.75 \\
$\beta$ = 58.2     &  F3  &  0.25 &   0.65 &  0.84 \\
$\gamma$ = 58.2     &  F4 &  0.15 &   0.75 &  0.34 \\
V = 100.9         &  F5   &  0.84 &   0.25 &  0.65 \\
~                 &  F6   &  0.75 &   0.34 &  0.15 \\
\hline
\hline
\end{tabular}
}
\hfill
\parbox{.45\linewidth}{
\centering
\begin{tabular}{c c c c c}
\hline
\hline
Lattice & Atoms & $x$ & $y$ & $z$ \\
\hline
a = 5.0806          &  Cu1  &  0.50 &   0.50 &  0.50 \\
b = 5.0806          &  Cu2  &  0.00 &   0.00 &  0.00 \\
c = 5.0806          &  F1   &  0.75 &   0.10 &  0.39 \\
$\alpha$ = 56.65$^\circ$    &  F2  &  0.34 & 0.15 & 0.75 \\
$\beta$ = 56.65$^\circ$     &  F3  &  0.10 & 0.39 & 0.75 \\
$\gamma$ = 56.65$^\circ$    &  F4  &  0.60 & 0.25 & 0.89 \\
V = 85.55~\AA$^{3}$         &  F5  &  0.39 & 0.75 & 0.10 \\
~                           &  F6  &  0.89 & 0.60 & 0.25 \\
\hline
\hline
\end{tabular}
}
\end{table}

\begin{table}[hb!]
\caption{Total energy in eV per formula unit obtained by DFT.}
\label{DFTU_toten}
\begin{center}
\begin{tabular}{c c c c c}
\hline
\hline
DFT & $R\bar3c$-CuF$_3$ & $Pnma$-CuF$_3$ & Cu$_2$F$_5$ & $Im\bar3m$-K$_{0.75}$CuF$_3$ \\
\hline
Spin-unpolarized  & -29.2752 & -29.2414  & -26.7237  & -155.4913  \\
Spin-polarized & -29.4912  & -29.4518  & -26.8182 & -155.4913  \\
\hline
\hline
\end{tabular}
\end{center}
\end{table}

\begin{table}[hb!]
\caption{Total energy in eV per formula unit obtained by DFT+U.}
\label{DFTU_toten}
\begin{center}
\begin{tabular}{c c c c c}
\hline
\hline
Magnetic order & $Bmab$-La$_2$CuO$_4$ & Cu$_2$F$_5$ & $Pnma$-CuF$_3$ & $R\bar3c$-CuF$_3$ \\
\hline
FM  & -49.4917 & -22.8729  & -12.5685  & -12.2429  \\
AFM & -49.7076 & -23.2520  & -12.6731  & -12.6993  \\
\hline
\hline
\end{tabular}
\end{center}
\end{table}

\begin{figure}[h!]
\begin{minipage}[h]{0.32\linewidth}
\center{\includegraphics[width=1\linewidth]{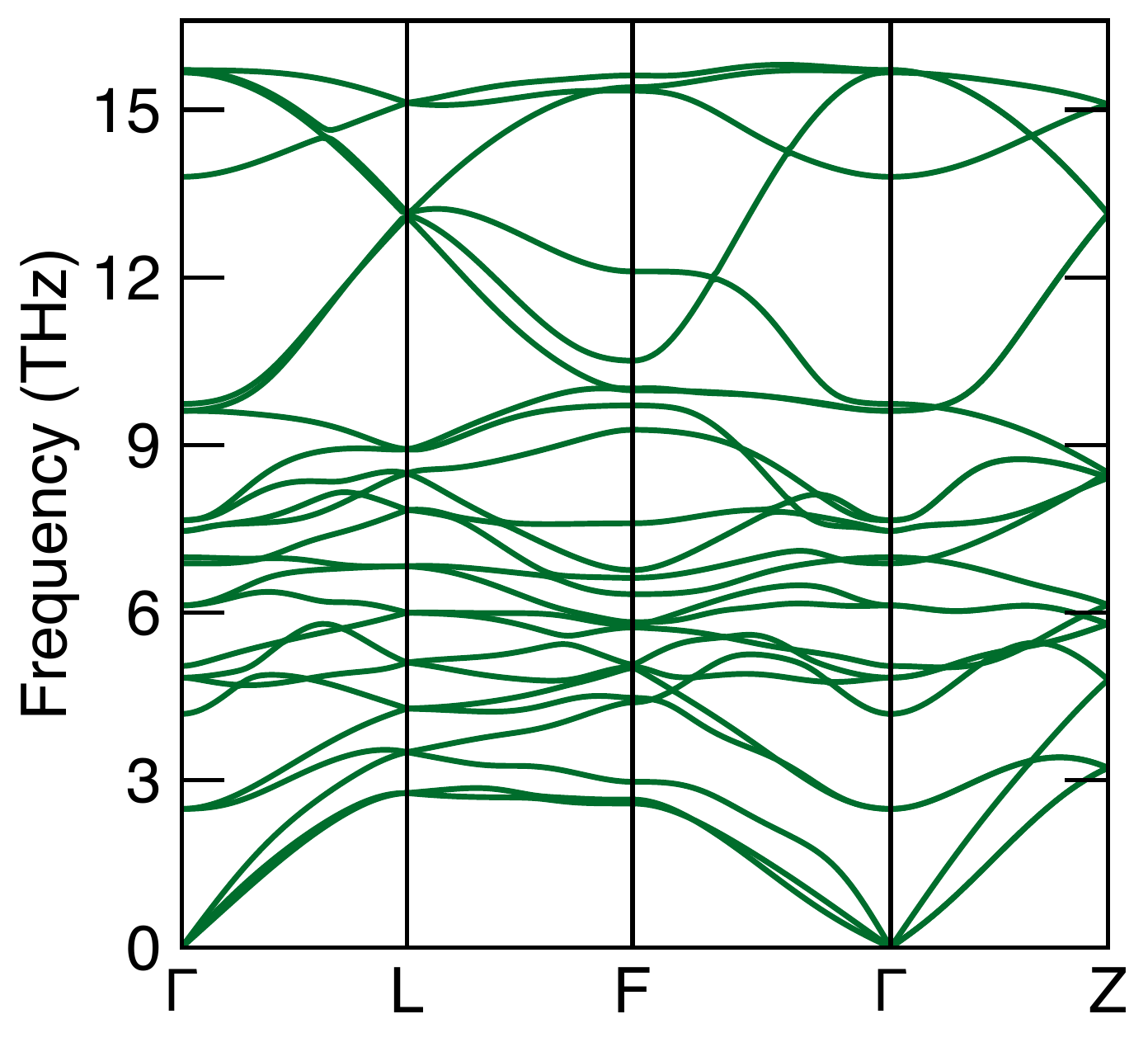}} (a) \\
\end{minipage}
\hfill
\begin{minipage}[h]{0.32\linewidth}
\center{\includegraphics[width=1\linewidth]{cuf3_r-3c_ph.pdf}} (b) \\
\end{minipage}
\begin{minipage}[h]{0.32\linewidth}
\center{\includegraphics[width=1\linewidth]{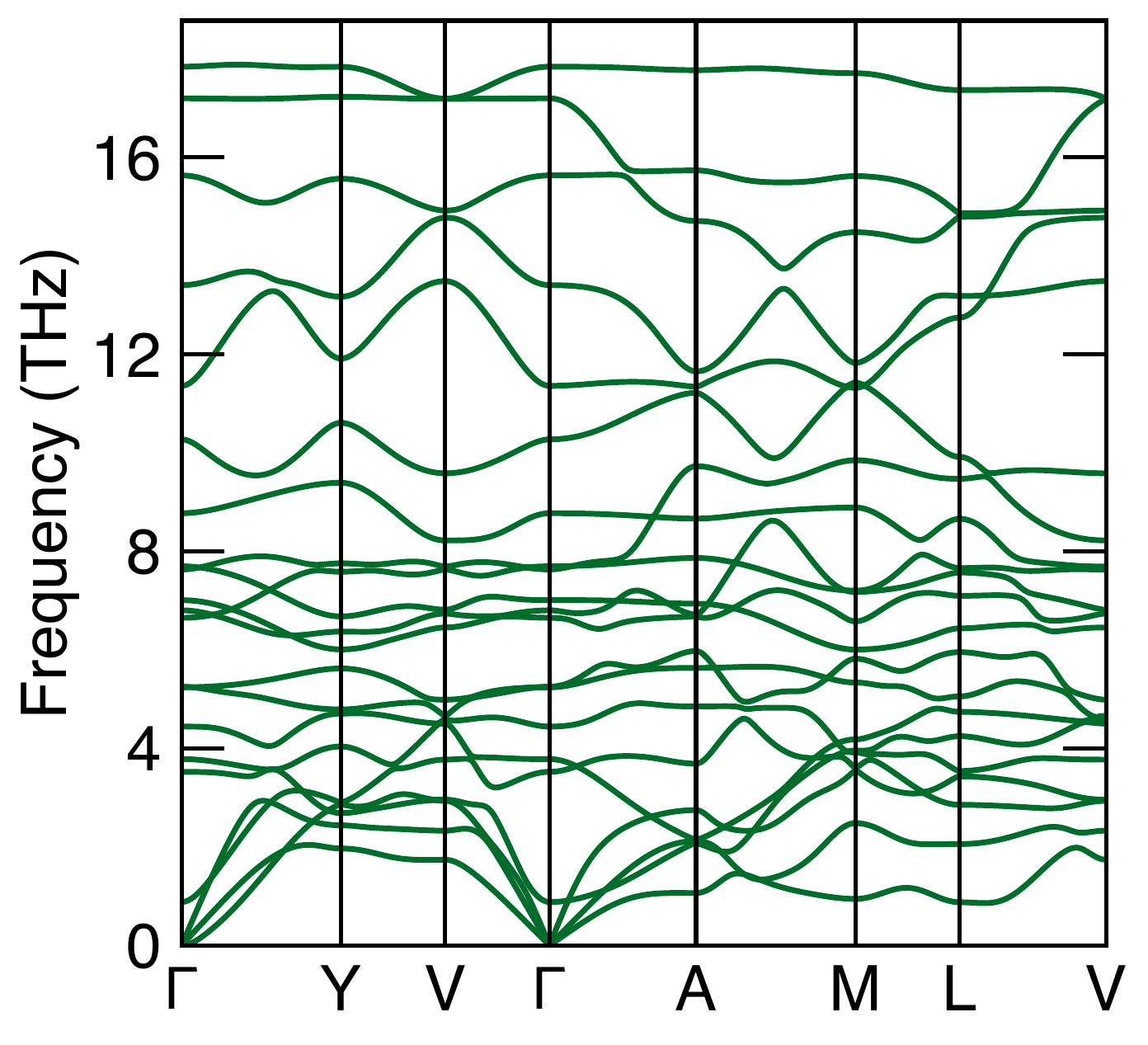}} (c) \\
\end{minipage}
\hfill

\caption {Phonon spectrum of (a) $Pnma$-CuF$_3$, (b) $R\bar3c$-CuF$_3$, and (c) $C2/m$-Cu$_2$F$_5$. We took supercell constructed from the primitive cell: $3\times3\times3$ for $R\bar3c$-CuF$_3$ and $C2/m$-Cu$_2$F$_5$, and $2\times2\times3$ for $Pnma$-CuF$_3$.} 
\label{fig:pdos_of_CuF3_r3c_and_La2CuO4}
\end{figure}

\begin{figure}[h!]
\begin{minipage}[h]{0.49\linewidth}
\center{\includegraphics[width=1\linewidth]{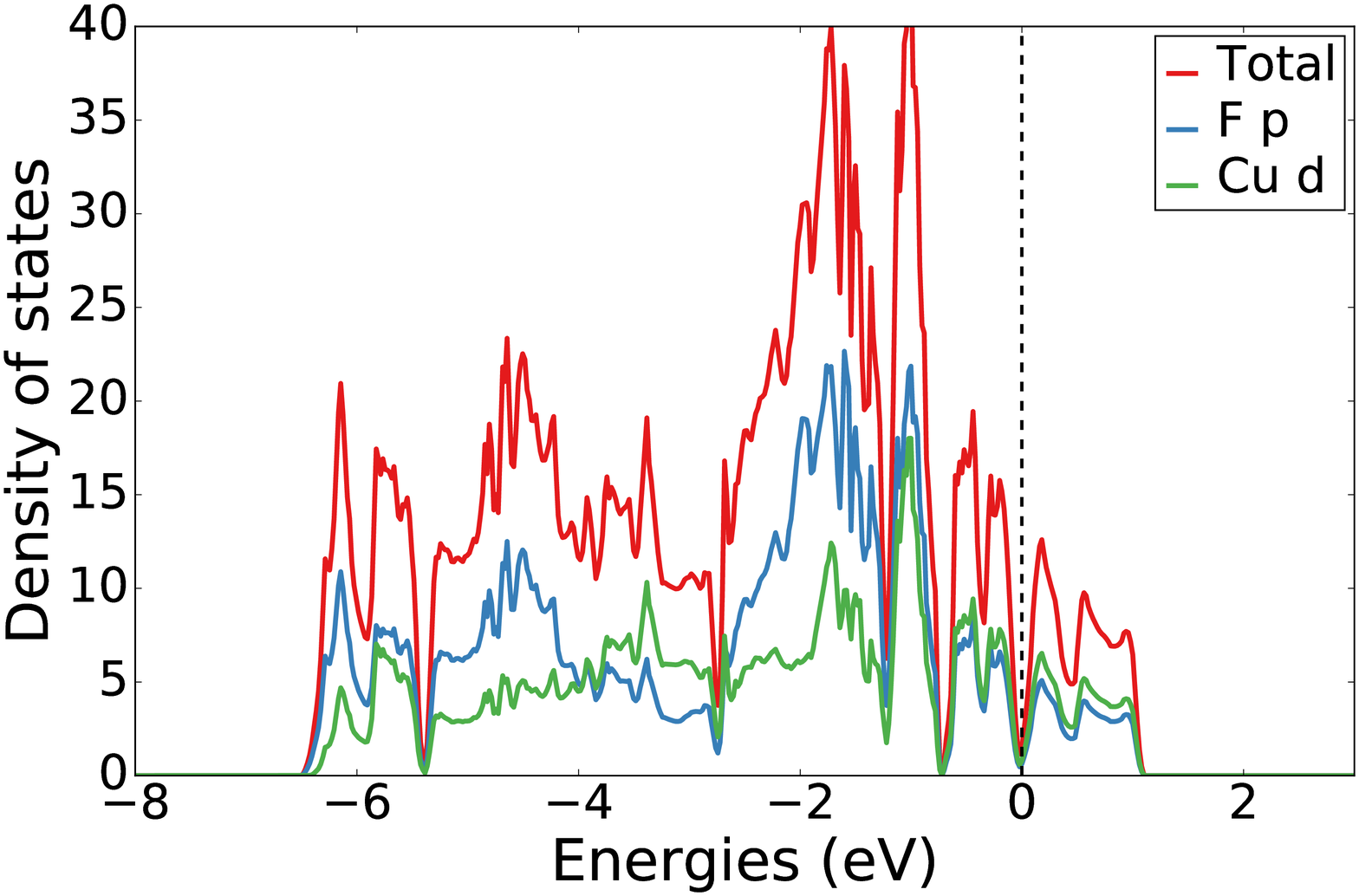}} (a) \\
\end{minipage}
\hfill
\begin{minipage}[h]{0.49\linewidth}
\center{\includegraphics[width=1\linewidth]{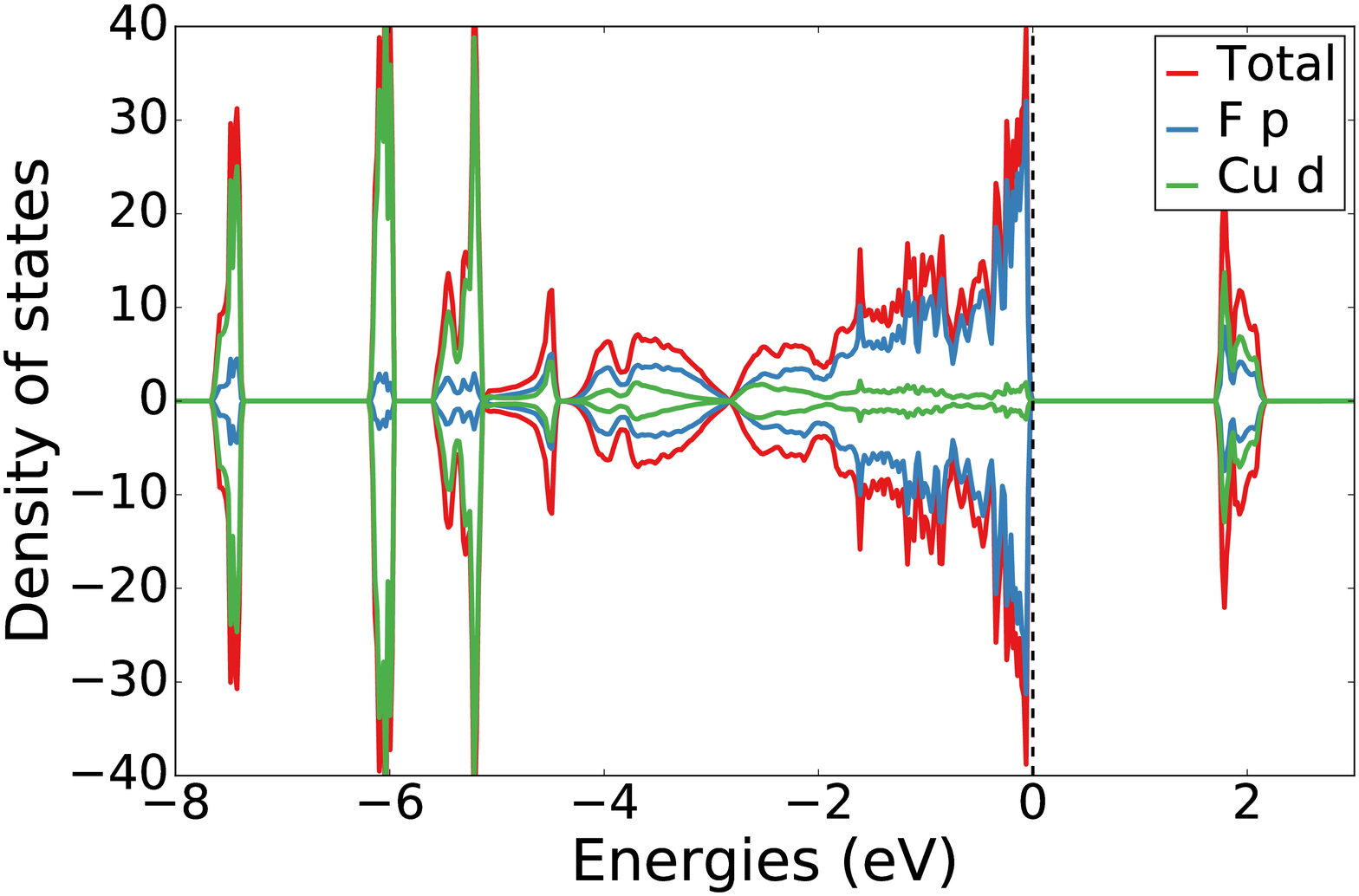}} (b) \\
\end{minipage}
\caption{Density of states of $Pnma$-CuF$_3$ obtained using (a) DFT and (b) DFT+U.}
\label{fig:CuF3_Pnma}
\end{figure}

\begin{figure}[h!]
\begin{minipage}[h]{1\linewidth}
\center{\includegraphics[width=0.8\linewidth]{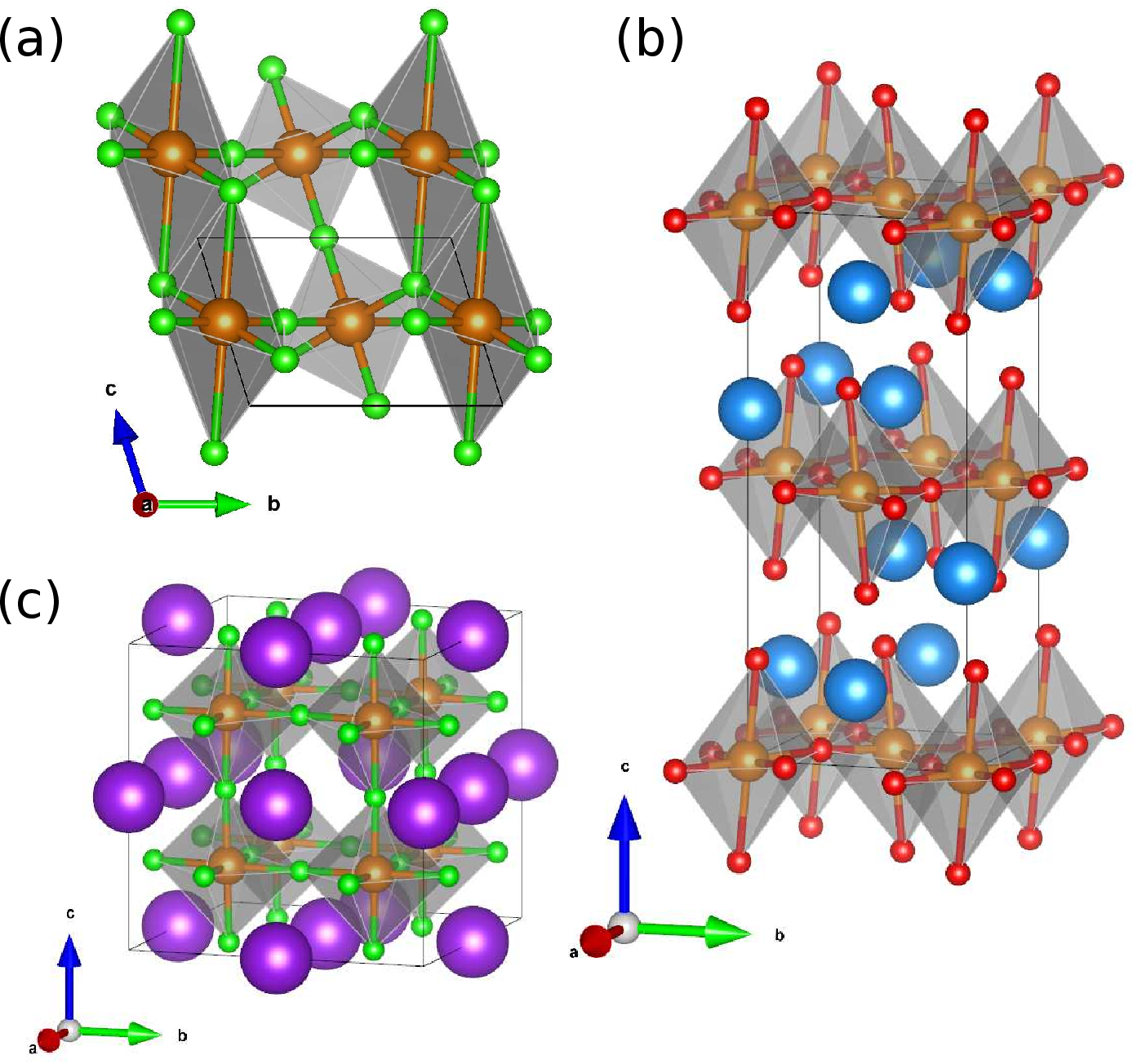}}
\end{minipage}
\caption {(color online) Schematic representation of the crystal structures: (a)~$C2/m$-Cu$_2$F$_5$,(b)~$Pmab$-La$_2$CuO$_4$, and (c)~$Im\bar3m$-K$_{0.75}$CuF$_3$. The Cu atom is brown, F is green, O is red, K is purple, and La is blue.}
\label{fig:cuf_ch}
\end{figure}
\appendix
\printbibliography